\documentstyle[12pt,a4,epsfig,a4wide]{article} 

\def\Jo#1#2#3#4{{#1} {\bf #2}, #3 (#4)}

\def\NPB{{Nucl. Phys.} {\bf B}}
\def\NPBP{{Nucl. Phys.} {\bf B} (Proc. Suppl.)}

\def\PLB{{Phys. Lett.}  {\bf B}}
\def\PRL{Phys. Rev. Lett.}
\def\PRD{{Phys. Rev.} {\bf D}}

\def\PRP{{Phys. Rep. }}
\def\EPC{{Eur. Phys. J.} {\bf C}}

\def\PNPP{Prog. Nucl. Part. Phys.}
\def\PTP{Prog. Theo. Phys.}
\def\PTPS{Prog. Theo. Phys. Suppl.}

\def\RMP{Rev. Mod. Phys.}
\def\MPL{Mod. Phys. Lett. {\bf A}}

\def\JHEP{J. High Energy Phys.}

\def\be{\begin{equation}}
\def\ee{\end{equation}}
\def\gs{\mathrel{
   \rlap{\raise 0.511ex \hbox{$>$}}{\lower 0.511ex \hbox{$\sim$}}}}
\def\ls{\mathrel{
   \rlap{\raise 0.511ex \hbox{$<$}}{\lower 0.511ex \hbox{$\sim$}}}}
\newcommand{\obb}{0\mbox{$\nu\beta\beta$}}
\newcommand{\onbb}{neutrinoless double beta decay}

\newcommand{\ba}{\begin{array}{c}}
\newcommand{\baz}{\begin{array}{cc}}
\newcommand{\bad}{\begin{array}{ccc}}
\newcommand{\bea}{\begin{equation} \begin{array}{c}}
\newcommand{\eea}{ \end{array} \end{equation}}
\newcommand{\ea}{\end{array}}
\newcommand{\D}{\displaystyle}

\newcommand{\aen}{\mbox{$\overline{\nu}_e$ }}

\newcommand{\nual}{\mbox{$\nu_{\alpha L}$ }}

\newcommand{\nualc}{\mbox{$(\nu_{\alpha L})^c$ }}
\newcommand{\nubec}{\mbox{$\nu_{\beta R}^c$ }}
\newcommand{\anual}{\mbox{$\overline{\nu_{\alpha L}}$ }}

\newcommand{\meff}{\mbox{$\langle m \rangle$ }}

\newcommand{\mab}{\mbox{$\langle m_{\alpha \beta} \rangle $}}

\newcommand{\mee}{\mbox{$\langle m \rangle $}}

\newcommand{\tm}{\mbox{$\tilde{m}$}}
\begin{document}
\newpage
\title{\hfill { \bf {\small DO--TH 00/12}}\\
\hfill { \bf {\small hep-ph/0008044}}\\ \vskip 1cm  
{\bf Cancellations in Neutrinoless Double Beta Decay and the 
Neutrino Mass Matrix}}
\author{Werner Rodejohann\footnote{Email address: rodejoha@xena.physik.uni-dortmund.de}\\
{\it \normalsize Lehrstuhl f\"ur Theoretische Physik III,}\\ 
{\it \normalsize Universit\"at Dortmund, Otto--Hahn Str. 4,}\\ 
{\it \normalsize 44221 Dortmund, Germany}}
\date{}
\maketitle
\thispagestyle{empty}
\begin{abstract}
In a degenerate scheme with mass $m_0$ 
a complete analysis of the allowed 
range of the effective electron neutrino Majorana mass 
$\langle m \rangle$ is performed. 
Special attention is paid to effects of cancellations caused either by 
intrinsic $CP$ parities of the eigenstates ($CP$ invariance) or 
by complex mixing matrix elements ($CP$ violation). 
We investigate all possibilities 
and give in each case constraints on the phases, the 
relative $CP$ parities or the neutrino mass scale. 
A solar mixing angle $\sin^2 2 \theta$ smaller 
than 0.7 jeopardizes the degenerate mass scheme. 
A key value of $\langle m \rangle/m_0$ is identified, which 
is independent on the solar solution and 
would rule out certain schemes. Also it would answer 
the question regarding the presence of 
$CP$ violation.    
Even if a total neutrino mass scale and 
an effective mass is measured, the value of the phases or parities 
is not fixed, unless in some special cases. 
The resulting uncertainty in the other mass matrix elements is 
at least of the same order than the one stemming from 
nuclear matrix elements calculations.

\end{abstract}

\newpage
\section{\label{eins}Introduction}
In the light of recent impressive experimental evidence on 
neutrino oscillations \cite{reviews} the next fundamental question to 
be answered is the one regarding the neutrino character. 
From the 
theoretical side Majorana neutrinos are favored, since 
they pop out of almost every GUT and are for example the result of the 
very attractive see--saw--mechanism \cite{seesaw}. In this case, heavy 
neutrinos are predicted.  
Experimental information on the existence of Majorana neutrinos  
might come from \onbb{} (\obb{}) \cite{obb} 
or from 
production of heavy Majorana neutrinos at colliders, see e.g.\ 
\cite{iso} and references therein. The most stringent limit however 
comes from the first process, investigating the effective Majorana mass 
of the electron neutrino, with 
a current bound of  \cite{baudis}
\be \label{meelim}
\mee \ls 0.2 \;  {\rm  eV}. 
\ee
Plans exist to build experiments exploring regions 
up to $\mee \simeq 0.002$ eV \cite{genius}. 
The variation within a factor of roughly 3 between different calculations 
of the required nuclear matrix elements has to be kept in mind. 
Results of oscillation experiments can be used to restrict the value 
of \mee{} in different mass schemes and for the different solutions of 
the solar neutrino problem 
\cite{meebil,minyas,jap1,inder,glzr1,bar,viss,ich,jap2,klap,glzr2}. 
Typical key 
scales for \mee{} are 0.1 and 0.005 eV\@, thus lying in the range of 
current and forthcoming projects. 
However, \mee{} has a form which 
includes possibilities for the contributions to cancel each other, 
namely via $CP$--violating phases or via the intrinsic 
$CP$--parities of the mass states, which exist 
in the case of $CP$ conservation. 
These effects were included in most of 
the above given references to get the maximal and 
minimal values of \mee. 
In \cite{jap2} 
constraints on $CP$--violating phases were given by using a 
graphical representation for the complex mass matrix elements, although 
without applying numerical values.  
Numerical studies regarding the 
Majorana phases were given in \cite{minyas,inder}. 
Some overlap with these works exists, however, 
we give many plots and statements, which are new to the literature and adopt 
an approach, which allows to investigate different situations regarding 
measurements of \mee{} and/or $m_0$. Here $m_0$ denotes the common 
mass scale in a degenerate mass scheme.  
Among the topics discussed are a clarification of 
the kind of cancellation, i.e.\ a distinction between the 2 possibilities. 
When starting from a degenerate scheme and in a 
given situation the maximal allowed $m_0$ is comparable to the 
scale implied by the atmospheric anomaly, so the 
degenerate scheme is 
ruled out and the derived mass limit holds for the largest of the 
3 mass eigenstates. 
We will find that only in very special cases definite statements about the 
phases or parities can be made and that alone from this fact 
there is a large 
uncertainty in the values of the other mass matrix entries. 
This uncertainty is ranging from 2 to factors around 20. 
The other entries are clearly needed to distinguish between 
different discussed models. The only information about the phases or 
parities can come from \onbb:  
All other mass matrix entries are impossible to measure directly, since 
the respective branching ratios or cross sections they govern are 
way beyond experimental access \cite{ich}. 
Apart from the obvious aspect that fundamental parameters of 
a model need to be measured, 
other interesting application of the phases exist: 
In \cite{stab} it was found that the Majorana 
phases play a crucial role in the stability of the mixing angles 
against quantum corrections. 
Ref.\ \cite{joseap} finds that their values have influence on 
the magnitude of the lepton--asymmetry in the universe, 
which can be made responsible for the observed baryon--asymmetry. 
Thus, the precise knowledge of all mixing parameters in the lepton sector is 
certainly important.\\ 
The paper is organized as follows: In Section \ref{zwei} we present the 
general framework and basic formulae for $CP$ violation and conservation, 
respectively. Section \ref{drei} sees a discussion of 
the connection between oscillation 
and \mee{} in hierarchical schemes, whereas Sec.\ \ref{vier} 
concentrates on the degenerate scheme. Some special mixing matrices 
and the general case with experimentally favored values are 
discussed. 
For each case constraints on the phases, parities and $m_0$ are given. 
The paper 
is concluded in Sec.\ \ref{funf} with a summary of our results.

\section{\label{zwei}Formalism}
Flavor eigenstates $\nu_{\alpha}$ 
($\alpha = e ,\, \mu , \, \tau$) 
are connected to mass eigenstates $\nu_i$ ($i = 1,2,3$) 
via a mixing matrix, i.e.\ $\nu_{\alpha} = S_{\alpha i} \nu_i$. 
A proper treatment of this issue can be found in \cite{proper}, here 
we quote only the main points. 
This matrix is derived by diagonalising the Majorana mass term in the 
Langrangian: 
\be \label{massterm}
- \mbox{$\cal L$}_{\rm mass} = \frac{1}{2} 
M_{\alpha \beta} \anual \nubec + \rm h.c. 
\ee
where $U_{CP} \nual U_{CP}^{-1} = \nu_{\alpha R}^c$ 
and $M_{\alpha \beta}$ is a symmetric matrix. For $CP$ invariance 
it is also real and can be diagonalized by an orthogonal matrix 
\be \label{O}
O_{i \alpha} M_{\alpha \beta} O_{\beta j}^T = \eta_i m_i \delta_{ij}
\ee
with $\eta_i = \pm 1$ and $m_i \ge 0$. Choosing 
$\chi_i = \sum_a O_{i \alpha} \nual{} + \eta_i O_{i \alpha} \nualc{}$ 
we have $- 2 \mbox{$\cal L$}_{\rm mass} = m_i \overline{\chi}_i \chi_i$ and 
$U_{CP} \chi_i U_{CP}^{-1} = \eta_i \chi_i$. 
For $CP$ invariance it can be shown \cite{bilpet} that 
$\eta_i$ is connected to the intrinsic $CP$ parity of the Majorana, 
i.e.\ $\eta_i = i \eta^{CP}_i $. In addition it holds 
$\chi^c = \pm \eta_i \chi$.  
Therefore the \obb{} amplitude for $CP$ invariance is proportional 
to \cite{kayser}
\be \label{MCPI}
\mbox{$\cal M$}_{CP}(\obb) \propto O_{ei}^2 \gamma_- 
\chi_{i} \overline{\chi_{i}^c} \gamma_- 
\propto \eta_i m_i O_{ei}^2 , 
\ee
with $2 \gamma_\pm = 1 \pm \gamma_5$. 
At the cost of a complex $O$ 
the $CP$ parities can also be absorbed in the mixing matrix via the identity 
$\eta_i = e^{ i \pi/2 (\eta_i - 1)}$. 
For complex  $M_{\alpha \beta}$ 
and thus $CP$ violation\footnote{A geometrical description of $CP$ violation 
with Majorana neutrinos in terms of unitarity 
triangles can be found in \cite{geocp}.} 
the diagonalization of the mass term 
is done by an unitary matrix, 
\be \label{U}
U_{i \alpha} M_{\alpha \beta} U_{\beta j}^T = m_i \delta_{ij}. 
\ee
Appropriate choice of $U$ can always make $m_i \ge 0$.  
Here, $\eta_i = 1$ and the \obb{} amplitude is 
proportional to 
\be \label{MCPV}
\mbox{$\cal M$}_{CP\hspace{-.6em}{\mbox{/}}}
(\obb) \propto U_{ei}^2 \gamma_- 
\chi_{i} \overline{\chi_{i}^c} \gamma_- 
\propto m_i U_{ei}^2 . 
\ee
To conclude, the matrix in the relation $\nu_{\alpha} = S_{\alpha i} \nu_i$ 
is either orthogonal or unitarity and the 
quantity probed in \obb{} is 
\be \label{conclude} 
\mee = \left\{ \baz |\D \sum_i U_{ei}^2 m_i |& CP \mbox{ violation} \\[0.3cm]
                    |\D \sum_i O_{ei}^2 m_i \eta_i | & CP \mbox{ invariance}
 \ea \right. . 
\ee  
For 2 flavors the two cases read: 
\be \label{2flav} 
\mee = \left\{ \baz |m_1 \cos^2 \theta  + m_2  \sin^2 \theta e^{2 i \phi} | 
                    & CP \mbox{ violation} \\[0.2cm]
                    |m_1 \cos^2 \theta + m_2 \eta_1 \eta_2 \sin^2 \theta | 
                    & CP \mbox{ invariance}
 \ea \right. .
\ee
Hence, a single $CP$ parity has no physical meaning, only relative values are 
significant. Note though that 
the case $\eta_1 \eta_2 = -1$ (opposite parities) 
and $\phi = \pi/2$ (``maximal'' violation) can not be distinguished. 
This can be also seen from the choice of the mixing matrix $S_{\alpha i}$: 
The case $\eta_1 = -1$ and $S_{e1} = O_{e1}$ is equivalent to 
$\eta_1 = +1$ and $S_{e1} = e^{-i \pi/2} O_{e1} = -i O_{e1}$. This means 
that opposite parities with a real mixing matrix are equivalent to 
equal parities with a complex mixing matrix for a maximal phase. 
Hence, for ``maximal'' phases we can not tell from \obb{} alone if 
there is $CP$ violation in the lepton sector and the answer if there is one 
at all has to come from long--baseline 
experiments \cite{lbl}. 
This is similar to the statement by Pal and Wolfenstein \cite{wolfpal} 
for the decay $\nu_2 \to \nu_1 \gamma$: 
For equal (opposite) parities magnetic 
(electric) dipole radiation occurs. Opposite parities are equivalent to 
complex mixing matrix elements but do not imply $CP$ violation. 
This would be suggested only by the presence of both kinds of 
radiation. For a degenerate scheme with masses discussed in 
the following sections however, the 
life time of such a process is about $\tau \gs 10^{48}$ s.

\section{\label{drei}Oscillation Experiments, \obb{} and \mee{} 
in Hierarchical Schemes}
The $CP$--violating mixing matrix $U$ can be parametrized as 
\bea \label{Upara}
U = U_{\rm CKM} \; 
{\rm diag}(1, e^{i \alpha}, e^{i (\beta + \delta)}) \\[0.3cm]
= \left( \bad 
c_1 c_3 & s_1 c_3 & s_3 e^{-i \delta} \\[0.2cm] 
-s_1 c_2 - c_1 s_2 s_3 e^{i \delta} 
& c_1 c_2 - s_1 s_2 s_3 e^{i \delta} 
& s_2 c_3 \\[0.2cm] 
s_1 s_2 - c_1 c_2 s_3 e^{i \delta} & 
- c_1 s_2 - s_1 c_2 s_3 e^{i \delta} 
& c_2 c_3\\ 
               \ea   \right) 
 {\rm diag}(1, e^{i \alpha}, e^{i (\beta + \delta)}) , 
\eea
where $c_i = \cos\theta_i$ and $s_i = \sin\theta_i$. 
The orthogonal matrix $O$ is of course obtained 
by setting the phases to zero. From hereon we will 
always write $U_{\alpha i}$ for the mixing matrix 
with obvious changes for the $CP$ conserving 
case. Since the oscillation 
probability 
\be \label{prop}
P_{\alpha \beta} = \delta_{\alpha \beta} - 2 \mbox{ Re } \sum\limits_{j > i} 
U_{\alpha i} U_{\alpha j}^{\ast} U_{\beta i}^{\ast} U_{\beta j} 
(1 - \exp{i \Delta_{ji}})  .
\ee
is invariant under such a multiplication of a diagonal matrix, 
the two additional Majorana induced phases are not observable 
in any oscillation experiment \cite{noph}. 
Therefore, \obb{} is the only 
probe to test these phases. In principle it would be possible to completely 
derive all phases by comparing the elements of the Majorana mass 
matrix 
\be
\mab = |(U \, {\rm diag}(m_1 \eta_1 , m_2 \eta_2 , m_3 \eta_3 ) 
U^{\rm T})_{\alpha \beta}| .
\ee
However, the analogues of \obb{} one needs to observe like 
$K^+ \to \pi^- \mu^+ \mu^+$ \cite{kaon}, 
$\nu_\mu N \to \mu^- \mu^+ \mu^+ X $ \cite{FRZ1} 
or $e^+ p \to \aen \mu^+ \tau^+ X$ \cite{FRZ2} 
have branching ratios or cross section far too small to be 
measured \cite{ich}.\\
All numerical analyses indicate a hierarchy in the mass squared differences 
of solar and atmospheric experiments: 
\be \label{m2hie} 
\Delta m^2_{\odot} \ll \Delta m^2_{A} . 
\ee
Maximal values are 
$(\Delta m^2_{\odot})_{\rm max} \simeq 10^{-4}$ eV$^2$ \cite{Inder,Spanier} 
and $(\Delta m^2_{A})_{\rm max} \simeq 10^{-2}$ eV$^2$ \cite{atm}. 
Without loss of generality we assume $m_3 \ge m_2 \ge m_1$. 
In a 3 flavor picture, 
three hierarchical schemes, where at least for 
one mass eigenstate $m_i^2 \simeq \Delta m^2$ holds, 
are capable of explaining the 
relation (\ref{m2hie}), they are 
called 
``completely hierarchical'' 
($m_3 \simeq \sqrt{\Delta m^2_A} \gg m_2 \simeq 
\sqrt{\Delta m^2_{\odot}} \gg m_1$), 
``partially hierarchical'' 
($m_3 \simeq \sqrt{\Delta m^2_A} \gg m_2 \simeq m_1$) and 
``inverse hierarchical'' 
($m_3 \simeq m_2 = \sqrt{\Delta m^2_A} \gg m_1$). 
With this notation, 
the CHOOZ result demands a small $|U_{e3}|$ in the first two cases and 
a small $|U_{e1}|$ in the last one.  
The latter fact stems from the condition that for $m_3 \ge m_2 \ge m_1$ 
the completely and partially hierarchical schemes demand 
$\Delta m_{21}^2 = \Delta m_{\odot}^2 \ll \Delta m_{31}^2 
\simeq \Delta m_{32}^2$, whereas the inverse hierarchical scheme has to 
be given by $\Delta m_{32}^2 = \Delta m_{\odot}^2 \ll \Delta m_{21}^2 
\simeq \Delta m_{31}^2$. 
With a good approximation we can use 2 flavor fits, then the solar 
mixing angle gives 
\be \label{ue12}
c_1^2,s_1^2  \simeq \frac{1}{2} 
\left(1 \pm \sqrt{1 - \sin^2 2 \theta_\odot}\right) . 
\ee
The data we use are taken from Refs.\ \cite{Inder,Spanier} and given in 
Table \ref{data}. As usual, SA (LA) denotes the small (large) angle 
MSW, LOW the MSW low mass and VO the vacuum solution.  
The MSW \cite{MSW} resonance condition demands $|U_{e1}| \ge |U_{e2}|$. 
However, 
for the vacuum solution this might not be the case, we come back later on 
that point. In the inverse hierarchical scheme the resonance requires 
$|U_{e3}| \ge |U_{e2}|$. 
The CHOOZ experiment \cite{chooz} 
gives unfortunately only a limit on 
 $|U_{e3}|$ (in the inverse hierarchical scheme on $|U_{e1}|$), 
depending on the atmospheric mass scale, it reads 
at 90 $\%$ C. L. 
\be \label{cholim}
|U_{e3}|^2 \ls 0.01 \ldots 0.15 \, \mbox{ for } \, 
\Delta m^2_A \simeq 10^{-3} \ldots 10^{-2} \; \rm eV^2 \; \cite{SK}.
\ee 
For typical best--fit values of few$\times 10^{-3}$ eV the bound is 
about $|U_{e3}|^2 \ls 0.05$.
Even for the maximal allowed values of the mass squared differences 
and mixing angles 
\mee{} is always below 0.2 eV: 
\bea \label{hie}
\mee{} \le m_1 |U_{e1}|^2 + m_2 |U_{e2}|^2 + m_3 |U_{e3}|^2 \\[0.2cm]  \le 
\left\{ \bad  
m_1 + |U_{e2}|^2 \sqrt{\Delta m^2_\odot} + |U_{e3}|^2 \sqrt{\Delta m^2_A}
& \le m_1 + 0.022 & \rm completely \; hierarchical \\[0.2cm]
(|U_{e1}|^2 + |U_{e2}|^2) \sqrt{\Delta m^2_\odot} 
+ |U_{e3}|^2 \sqrt{\Delta m^2_A} 
& \le 0.85 m_1 + 0.015 & \rm partially \; hierarchical \\[0.2cm]
|U_{e1}|^2 m_1 + (1 - |U_{e1}|^2) \sqrt{\Delta m^2_A} 
& \le m_1 + 0.085 & \rm inverse \; hierarchical \\[0.2cm] 
\ea \right. . 
\eea
Note that there can be a additional contribution of 
$m_1 < \sqrt{\Delta m^2_\odot}$, which can be traced back to the fact that 
only mass squared differences are measured. If an experimental sensitivity of 
$10^{-2}$ eV on \mee{} is achieved, this might be of importance. However, 
at the present time with $\mee \le 0.2$ eV, all values of the phases and the 
parities are allowed. In order to make more definite 
statements about 
the effects of cancellations, we therefore apply a degenerate scheme in which 
$m_3^2 \simeq m_2^2 \simeq m_1^2 \equiv m_0^2$. 
Per definition, it should 
hold $m_0^2 \gg \Delta m_{\rm max}^2 \simeq 10^{-2}$. 
When in the following a maximal allowed mass scale of 
$m_0 \simeq 0.2 \ldots 0.3$ eV is 
derived so does that mean that the mass scheme is not ``completely 
degenerate'' but ``slightly'' hierarchical, i.e.\ an intermediate 
situation occurs. For example, if $m_{0, \, \rm max} = 0.2$ eV, 
then another eigenstate has a mass of 
$(0.2 - \varepsilon)$ eV, where $\varepsilon$ is small. In order to get 
the atmospheric scale of $10^{-2}$ eV$^2$ one finds 
$\varepsilon \simeq 0.03$ eV and the two 
eigenstates differ by about 15 $\%$.  
Then our $m_{0,\, \rm max}$ is the limit on the 
largest mass eigenstate $m_3$ 
with $m_3 \simeq  \mbox{ or } > m_2$. A limit of $m_0 \simeq 0.1$ eV 
corresponds to a hierarchical scheme. However, due to the 
uncertainty in the value of \mee{} and $\Delta m_A^2$ a definite statement 
is somehow difficult to make. Using the best--fit point for the 
atmospheric scale means that the degenerate scheme fails for about 
0.1 eV\@.

\section{\label{vier}\mee{} in the Degenerate Scheme}
The upper bound on $m_0$ comes from the tritium spectrum, 
which limits \cite{mnue}
\be \label{trilim} 
\sum_i |U_{ei}^2| m_i^2 = m_0^2 \le (2.8 {\rm \; eV})^2 . 
\ee
Forthcoming and ongoing projects intend to push the bound below 
1 eV \cite{trifut}.  
In addition, cosmological observations might be interpreted in terms of 
a total neutrino mass of $\sum m_\nu \simeq $ few eV \cite{cosmo}. 
By measuring anisotropies in the cosmic microwave background, 
MAP and PLANCK may probe values down to $\sum m_\nu \sim 0.5$ eV 
\cite{MP}. 
The interesting function to investigate is 
$ \mee/m_0 \equiv \tilde{m}$, 
which is 
depending on 4 parameters, either two angles and 2 phases or 
two angles and two relative parities. 
In the $CP$--violating case \tm{} reads 
\be \label{CPI}
\frac{\mee}{m_0} \equiv \tilde{m} = c_3^2 \sqrt{c_1^4 + s_1^4 + t_3^4 
+ 2 (s_1^2 t_3^2 c_{2(\alpha - \beta)}  
+ c_1^2 (s_1^2 c_{2 \alpha} + t_3^2 c_{2 \beta} ))}
\ee
where $t_3 = \tan \theta_3$ and $c_{ 2\alpha} = \cos 2 \alpha$. 
For $CP$ invariance it can be written as 
\be \label{CPV}
\tm= c_3^2 (\eta_1 c_1^2 + \eta_2 s_1^2 + \eta_3 t_3^2) = 
\left\{ 
\bad 1                               & (+++) 
& \leftrightarrow \alpha = \beta = \pi \\[0.2cm]
      |c_3^2 (c_1^2 - s_1^2 - t_3^2)|  & (+--) 
& \leftrightarrow \alpha = \beta = \frac{\pi}{2}\\[0.2cm]
      |c_3^2 (c_1^2 - s_1^2 + t_3^2)|  & (-+-) 
& \leftrightarrow \alpha = \frac{\beta}{2} = \frac{\pi}{2}\\[0.2cm]
     \cos 2 \theta_3                 & (--+) 
& \leftrightarrow \alpha = 2 \beta = \pi
\ea \right. , 
\ee
where all 4 possible $(\pm \pm \pm)$ signatures with 
the corresponding $CP$--violating phases are given. In addition, 
there are ($c_1$--dependent) solutions 
for the phases, which give the same \tm{} as special parity 
configurations, see below.  
Note that 
for $ (--+)$ the value is independent on the solar solution. 
The $(+--)$ signature is in fact the minimal value for the general 
$CP$--violating case, provided that $|U_{e1}| \ge |U_{e2}|$. For 
the inverse situation, the $(-+-)$ signature gives the minimal \tm.\\
The different treatment of $CP$ parities and phases may seem somewhat 
artificial since the first option is a special case of the latter one. 
However, due to the maximal values certain configurations result in 
they deserve special attention.

\subsection{Some special mixing matrices}
Three intriguingly simple matrices have been widely discussed:
\begin{itemize}
\item Single maximal \cite{monomax}\\
Maximal atmospheric mixing and a vanishing angle in solar and 
the CHOOZ experiment, resulting in $|U_{e1}| = 1$.
\item Bimaximal \cite{bimax}\\
Both solar and atmospheric mixing is maximal, CHOOZ's angle is zero. 
Then $|U_{e1}| = |U_{e2}| = 1/\sqrt{2}$ and $|U_{e3}| = 0$. 
\item Trimaximal \cite{trimax}\\
All elements have the same magnitude and 
$|U_{e1}| = |U_{e2}| = |U_{e3}| = 1/\sqrt{3}$. The model 
gives a poor fit to the oscillation data. 
\end{itemize}
Then one gets for the $CP$--conserving case 
\be
\tm =  
\left\{ 
\baz  1,1,1           & (+++) \\[0.2cm]
      1,0,\frac{1}{3} & (+--) \\[0.2cm]
      1,0,\frac{1}{3} & (-+-) \\[0.2cm]
      1,1,\frac{1}{3} & (--+) 
\ea \right. , 
\ee
for single, bi-- and trimaximal mixing respectively. 
Single and bimaximal mixing are special cases of the mixing angles from 
Table \ref{data} and might serve as a model to get a feeling for 
the bounds one obtains. 
With the current limit on \mee{} we get for 
values of \tm{} = $1,\frac{1}{3}$: $m_0 \le 0.2, 0.6$. 
Obviously, for $\tm = 0$ it follows $\mee = 0$. 
If there is $CP$ violation we have 
\be
\tm=  
\left\{ \baz 1 & \mbox{ single} \\[0.2cm]
             c_\alpha & \mbox{ bi} \\[0.2cm] 
\frac{1}{3} \sqrt{3 + 2 (c_{2 \alpha} + c_{2 \beta} + c_{2(\alpha - \beta)})} 
&  \mbox{ tri}
\ea \right. . 
\ee
Single maximal mixing means $m_0 \le 0.2$ eV\@, unobtainable by 
currently planned experiments. For bimaximal mixing we get 
with the current limit and assuming $m_0 = 1$ eV\@: $\alpha \ls 78.5^0$. 
For the trimaximal case Fig.\ \ref{triCPVfig} shows \tm{} 
as a function of one phase for different values of the second phase. Note 
the constant value for $\beta = \pi/2$.

\subsection{General Limits and Bounds}
\begin{figure}[ht]
\setlength{\unitlength}{1cm}
\begin{center}
\epsfig{file=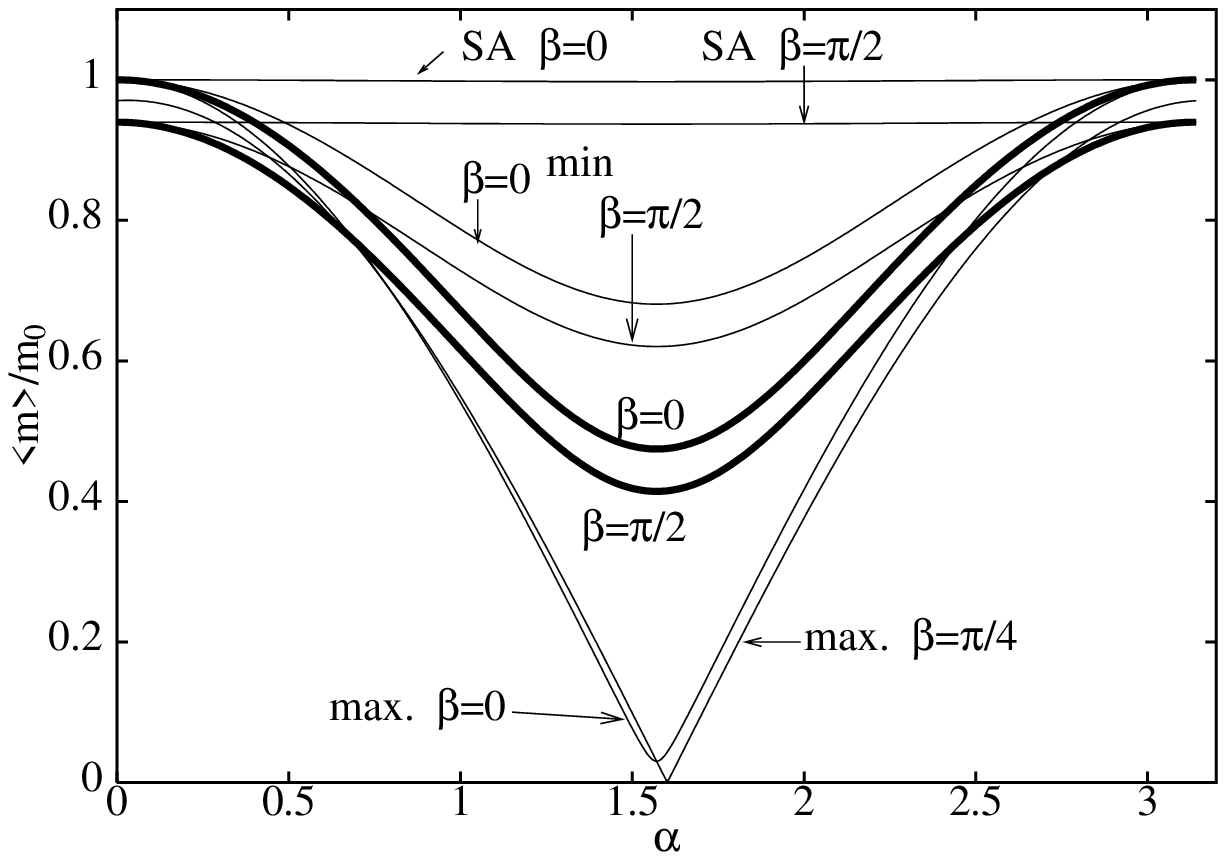,width=13cm,height=7.5cm}
\end{center}
\caption{\label{cpvfig1}\tm{} as a function of one phase $\alpha$ for 
different values of the second phase $\beta$. Displayed are the 
SA solution for $\beta=0$ and $\beta = \pi/2$ (the two straight lines) and 
the LA solution (best fit point are the thick lines) for the minimal allowed 
$sin^2  2 \theta_\odot$, maximal mixing and 
the best fit point (thick lines). 
Note that maximal mixing is also allowed for VO and LOW. 
We assumed $|U_{e3}|^2 = 0.03$.}
\begin{center}
\epsfig{file=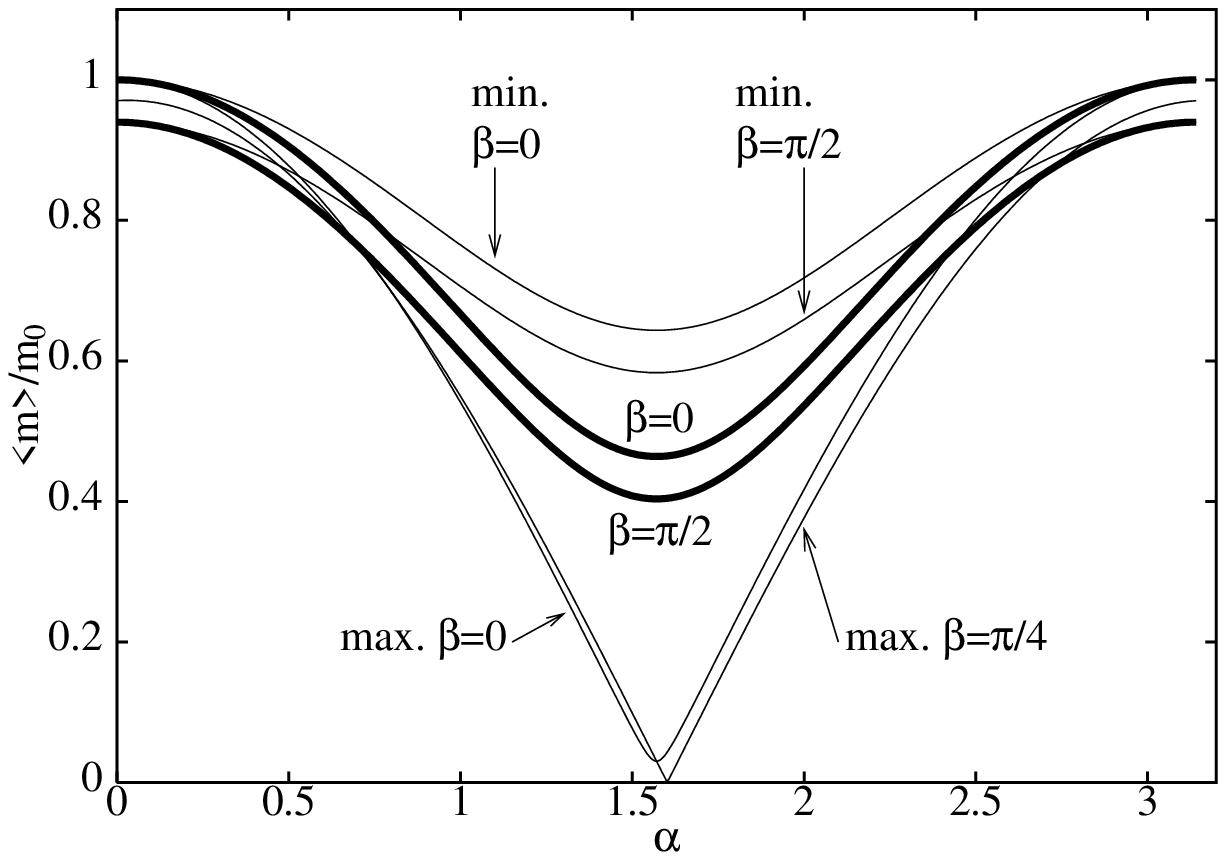,width=13cm,height=7.5cm}
\end{center}
\caption{\label{cpvfig2}Same as the previous figure for the VO solution.}
\end{figure}

\begin{figure}[ht]
\setlength{\unitlength}{1cm}
\begin{center}
\epsfig{file=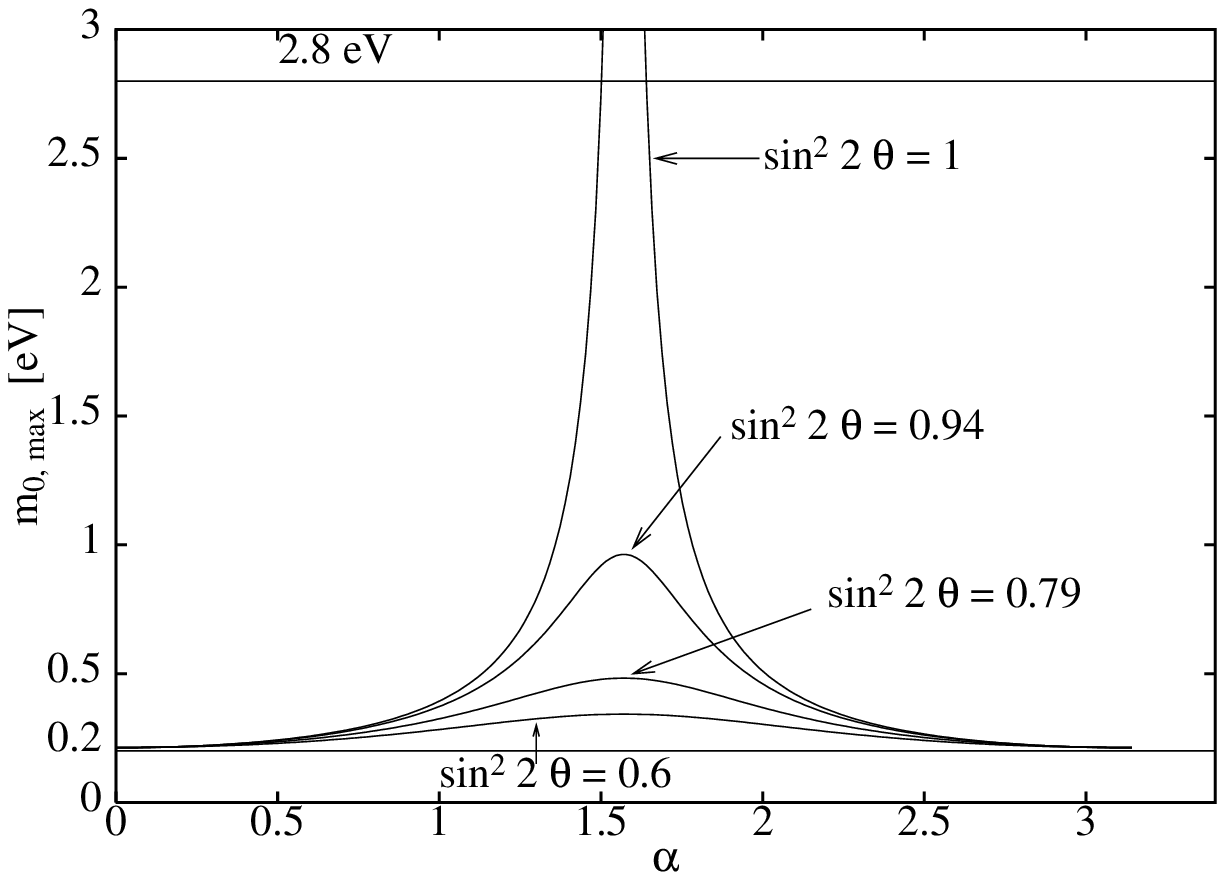,width=13cm,height=7.5cm}
\end{center}
\caption{\label{almofig}Maximal $m_0$ as a function of $\alpha$ for 
$\beta = \pi/2$ and $|U_{e3}|^2 = 0.03$ for different 
$\sin^2 2 \theta_\odot$. Allowed is the area under the respective curve. }
\begin{center}
\epsfig{file=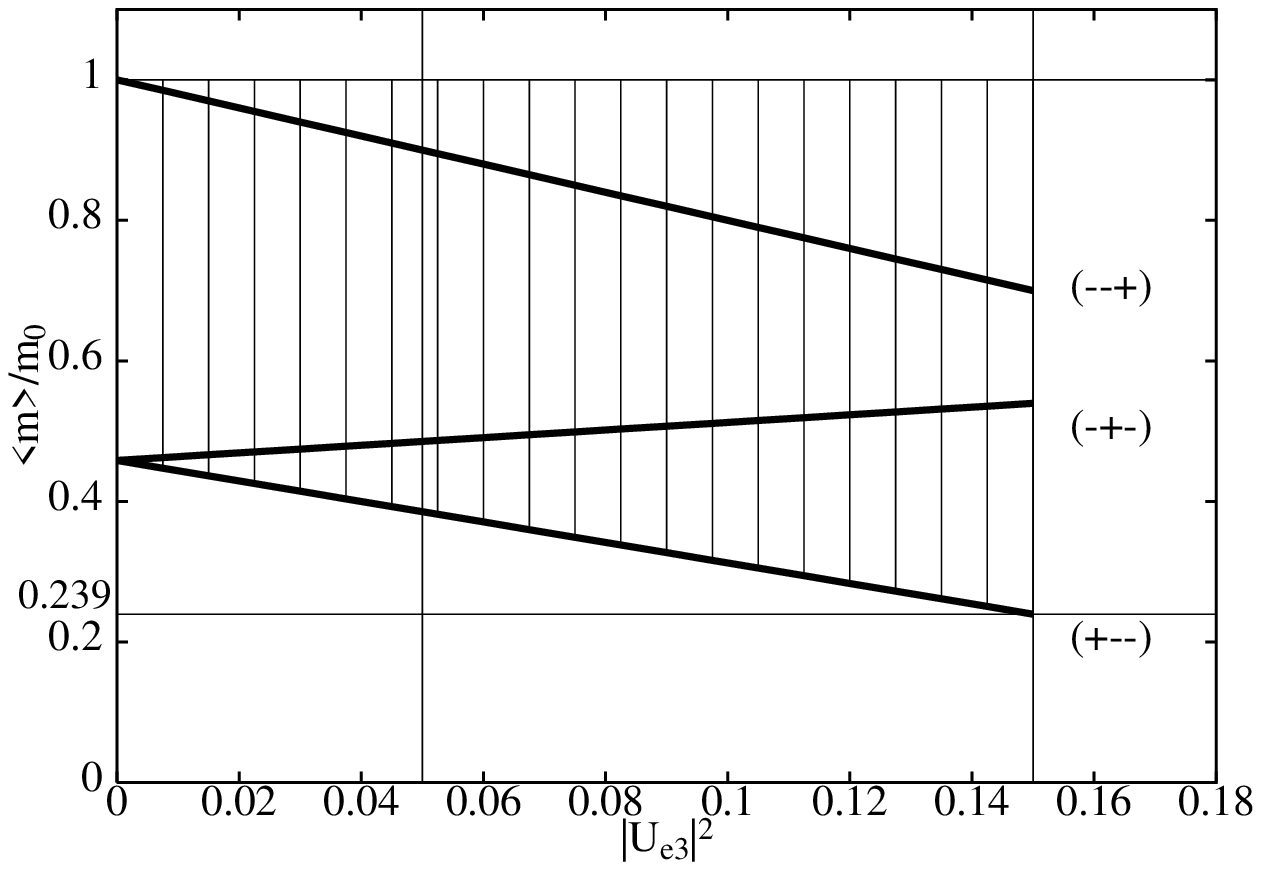,width=13cm,height=7.5cm}
\end{center}
\caption{\label{concl1}\tm{} as a function of $|U_{e3}|^2$ for 
the best--fit point of the LA solution. Displayed is the range 
for the $CP$--violating case and all $CP$--conserving parity configurations 
except the trivial $(+++)$ case. Note that the range of \tm{} 
accessible by currently planned experiments is $7 \cdot 10^{-4}$ to 0.4.}
\end{figure}

We go now from the special cases back to all experimentally 
allowed values of the mixing angels. The first thing to say is that 
for exactly vanishing $U_{e3}$ there is no way to find out about 
the second phase\footnote{Also, ``normal'' $CP$ violation in long--baseline 
experiments will be unobservable.}.
Without further effort we can say for the $(+++)$ signature, that 
$m_0 \le 0.2$, i.e.\ is beyond experimental access in cosmology or 
spectrum measurements. 
More interesting is e.g.\ the $(--+)$ case in which 
$\tilde{m}$ is depending on 
only one parameter, namely the angle bounded by CHOOZ\@. With Eqs.\ 
(\ref{meelim}) and (\ref{cholim}) 
we get the allowed range of $m_0$ depicted in Fig.\ \ref{mmpfig}. 
In this situation, the maximal neutrino mass is about 0.29 eV, 
using the best--fit point of SK gives $m_0 \le 0.22$ eV\@. We also 
plot the range for a limit of $\mee < 0.1$ eV\@, which further reduces 
the allowed values. The maximal $m_0$ is now 0.14 and 
0.11 eV\@, respectively. As can be seen, if cosmology insists in 
$m_0 \simeq 1$ eV, the $(--+)$ configuration is ruled out.\\
There is still 
freedom in the ordering of $|U_{e1}|$ and $|U_{e2}|$ if the 
vacuum solution is correct. We denote the choice 
$|U_{e1}| > |U_{e2}|$ with VO1 and the other one with 
VO2. However, as can be seen from Eqs.\ (\ref{ue12}) and (\ref{CPV}) 
the case VO1 and ($+--$) is equivalent to VO2 and $(-+-)$ as 
VO1 and ($-+-$) is to VO2 and ($+--$). 
In Figures \ref{pmmfig} and \ref{mpmfig} we show the maximal values 
of $m_0$ as a function of $|U_{e3}|^2$ for the current 
\mee{}--limit of 0.2 eV\@. These  
maximal values scale with the \mee{} limit. 
In \cite{inder} plots of the allowed  $|U_{e1}|^2$ and $|U_{e3}|^2$ were 
given for $m_0 = 1.7$ eV and all 3 nontrivial 
parity configurations. Their conclusion that small $|U_{e3}|$ requires 
near--maximal mixing for the $(-+-)$ and $(+--)$ case is consistent with 
Figs.\ \ref{pmmfig} and \ref{mpmfig}. For $(--+)$ they find that 
only small $m_0$ allows large mixing, which can be seen in 
Fig.\ \ref{mmpfig} as well. 
The authors also give the allowed areas in 
$|U_{e1}|^2$--$|U_{e3}|^2$ space for special values of the phases. 
However, to obtain these areas they included all $(\pm\pm\pm)$ 
possibilities, which is hard to compare with 
our approach. 
Reference \cite{minyas} also plots 
$|U_{e1}|^2$ against $|U_{e3}|^2$ for different situations resulting in 
similar conclusions as ours. Our plots are thus different projections of the 
5 dimensional parameter space giving complementary and additional 
information. Both mentioned works use cosmological arguments to set 
$m_0$ to 1.7 to 4 eV, whereas our approach (also used in \cite{glzr2}) 
allows to investigate 
different situations like positive results on \mee{} and/or $m_0$.\\
In Figs.\ \ref{cpvfig1} and \ref{cpvfig2} 
we plot \tm{} as a function of $\alpha$ for different 
values of $\beta$ whilst assuming $|U_{e3}|^2=0.03$. 
Maximal mixing (i.e.\ $|U_{e1}|=|U_{e2}|$) allows 
complete cancellation, whereas for other values a non--vanishing minimal 
\tm{} is achieved. The dependence on the second phase is rather 
small which can be explained by the smallness of $|U_{e3}|$, confer with 
Fig.\ \ref{triCPVfig}, where the dependence on the second phase is rather 
strong.  
In Fig.\ \ref{almofig} we therefore show for 
$\mee \le 0.2$ eV\@,  $\beta = \pi/2$ and 
$|U_{e3}|^2=0.03$ the allowed area in the $m_0$--$\alpha$ space 
for different $\sin^2 2 \theta_\odot$. The higher this angle is, 
the higher is the maximal allowed $m_0$. 
This value is approached for a solar mixing angle of about 0.7, which 
is also true for the $(+--)$ and $(-+-)$ configurations. 
For the SA solution it follows 
$c_1 \simeq 1$ in Eq.\ (\ref{CPV}) so that \tm{} is always about 1. 
This means if the SA solution turns out to be realized in nature direct 
mass searches will not find the neutrino 
mass scale because eutrinoless double beta decay then already demands 
$m_0 \le 0.2$ eV\@. 
If $m_0 \gs 0.5$ eV is measured and SA turns out to 
be true, then neutrinos are Dirac particles. These are well known facts 
and are only given for the sake of completeness.\\ 
The best--fit value and the minimal respectively maximal allowed 
mixing angle gives lines too close together 
to be distinguishable in the plot. 
Regarding VO1 and VO2 it turns out that the difference 
for given phases is negligible (1 to 2 $\%$), 
so we plot in Fig.\ \ref{cpvfig2} only 
the VO1 option. The difference between the two options would only be 
sizable for large $|U_{e3}|$.\\
If the bounds on \mee{} and $m_0$ are both further reduced, 
nothing special happens. 
Some more interesting possibilities however suggest themselves: 
First, a value for $m_0$ is found and \mee{} continues to give 
an upper bound.  
Then we get a {\it lower} limit on \tm. 
Second, the reverse situation results in an {\it upper} limit on \tm. 
If both quantities are measured, then definite statements about 
the phases can be made. 
In connection with the then already 
obtained knowledge of 
the correct solar solution and a more stringent CHOOZ bound (or signal) 
some important conclusions could be drawn. In addition, we can assume 
that the precision in the 
mixing angles will be improved. To simplify things a bit we take 
now the best--fit values. Since these 
points of the VO and the LA solution are very close together 
we only plot the LA and the LOW case in Figs.\ \ref{concl1} and 
\ref{concl2}, respectively. Recent data of the day--night spectrum 
seems to favor these solutions \cite{LAfav}. 
Different values of the parameters lead to modifications of the 
results which are not difficult to do with the help of 
Eqs.\ (\ref{ue12},\ref{CPI},\ref{CPV}). 
From the figures, we can easily analyze some specific configurations: 
For example, for $\tm{} < 0.6$ the $(--+)$ configuration is ruled out and 
for $\tm{} > 0.6$ the $(+--)$ and $(-+-)$ cases, making 0.6 a key scale, 
since it is independent on the solar solution and the value of $|U_{e3}|$. 
If we would know the value of \tm, even more could be said: 
The key value of 0.6 would mean that there is $CP$ violation 
in the leptonic sector. With the current 
limits on \mee and $m_0$ and the maximal sensitivities achievable 
by current 
experiments, only $\tm < 0.6$ seems realistic. The highest  
value with currently planned experiments is  
in fact $\tm = 0.2/0.5 = 0.4$, which for small $|U_{e3}|$ would  
rule out $(+--)$ and $(-+-)$ for LOW\@. Also, again for small 
$|U_{e3}|$, the LA and VO solution would be out of the game. 
Note however that the mentioned uncertainty for the nuclear 
matrix elements might allow to use a higher value for \meff, 
namely about $\sqrt{3}$\meff which then might 
allow to reach a \tm{} value of 0.6.\\
Conversely, the smallest number one can expect to measure is 
$\tm = 2 \cdot 10^{-3}/2.8 = 7 \cdot 10^{-4}$, which 
forbids the degenerate scheme at all and demands 
hierarchical or intermediate 
schemes. 
We refer to Refs.\ \cite{klap,glzr2} for the allowed values of \mee{} in 
that case. 
In general, as can be seen from the Figures,  
a value of \tm{} below 0.24 (0.06) rules out 
the degenerate scheme for the LA and VO (LOW) solution. 
Alternatively, if LA (LOW) turns out to be true and a \tm{} bound 
of 0.24 (0.06) is achieved, neutrinos are Dirac particles. 
Note however that our discussion relies on precise knowledge of 
the mixing angles and \tm{} and has thus to be taken with care.  
A realistic assumption about the minimal uncertainty in reach for 
these measurements might be given by 20 $\%$ for the former and a 
factor of 2 for the latter. 
We will show in the next section that the uncertainty for the other 
mass matrix elements is at least of this order.

\subsection{Consequences for the Mass Matrix}
\begin{figure}[ht]
\setlength{\unitlength}{1cm}
\begin{center}
\epsfig{file=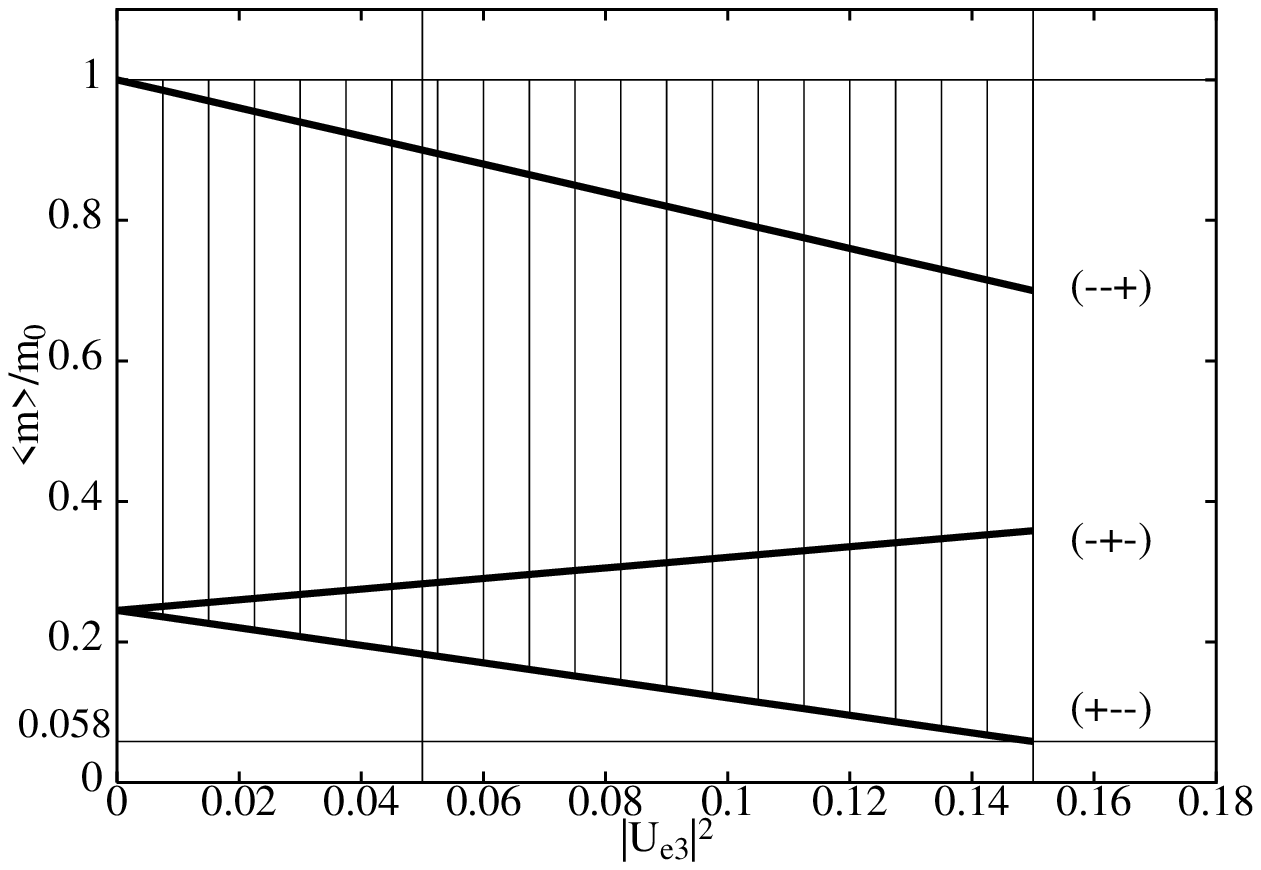,width=13cm,height=7.5cm}
\end{center}
\caption{\label{concl2}Same as the previous figure for the LOW solution.}
\begin{center}
\epsfig{file=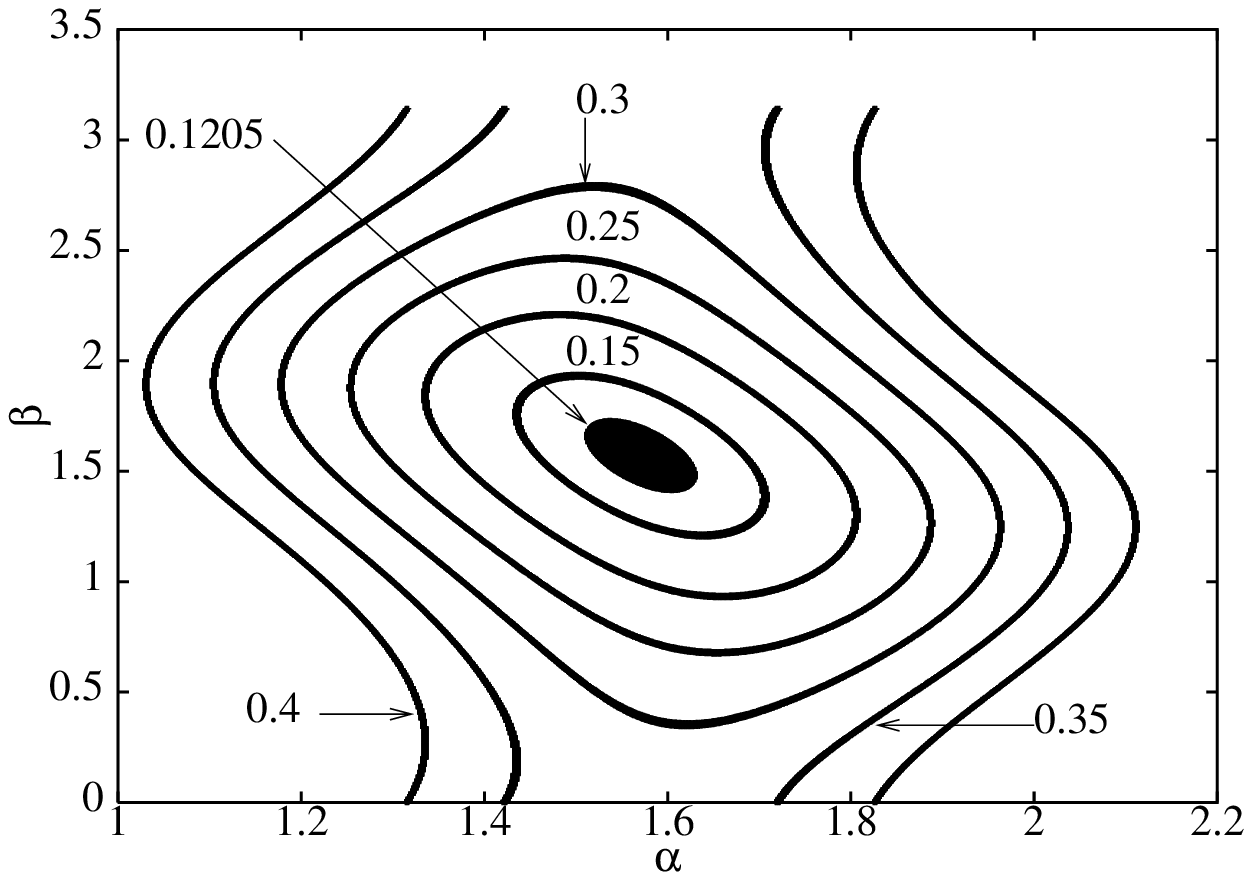,width=13cm,height=7.5cm}
\end{center}
\caption{\label{sametm}''Iso--\tm'' lines in the $\alpha$--$\beta$ 
space for the best--fit values of the LOW solution and $|U_{e3}|^2=0.1$. 
For \tm=0.1205 a precision of 5 $\%$ is assumed, the other values are for 
an exact measurement.}

\end{figure}

We want to show that the uncertainty in the other mass matrix entries is 
at least as large as the one stemming from $m_0$, nuclear matrix elements and 
the solar mixing angle. To show this we shall assume that both 
quantities as well as $|U_{e3}|$ and the angle governing the 
atmospheric neutrino anomaly are known precisely. 
As an example we take $\tm = 0.3$ and $|U_{e3}|^2 = 0.1$. 
This might be measured by $\mee = 0.18$ eV and $m_0 = 0.6$ eV\@. 
However, different choice of the 2 phases can result in the same \tm, as 
can be seen in Fig.\ \ref{sametm} for the LOW case, where 
``iso--\tm'' lines are displayed.  
For example, $(\alpha \simeq 1.177, \beta \simeq 1.864)$ or 
$(\alpha \simeq 1.7691, \beta \simeq 0.5225)$ are possible solutions for 
$\tm = 0.3$. 
The $\mu\mu$ entry of the mass matrix, 
$m_{\mu\mu} = m_0 |(U_{\mu 1}^2 + U_{\mu 2}^2 + U_{\mu 3}^2)|$  
is still a function of the third phase $\delta$. 
Assuming maximal atmospheric mixing, the two results for $m_{\mu\mu}$ 
differ by a factor of 4 (for $\delta=0$) to 8 for ($\delta \simeq 2.8$). 
It turns out that for a given $\delta$ some specific choices of the 
other two phases give for the resulting $m_{\mu\mu}$ 
relative differences of up to 15. Thus, to verify the real solution, 
one would need a measurement of $m_{\mu\mu}$. 
The charged kaon decay $K^+ \to \pi^- \mu^+ \mu^+$, which is depending 
on $m_{\mu\mu}$ as \obb{} is on \mee, has a branching ratio given 
by 
\be
\frac{\Gamma (K^+ \to \pi^- \mu^+ \mu^+)}{\Gamma (K^+ \to {\rm all})} \simeq 
10^{-31} \left(\frac{m_{\mu\mu}}{{\rm eV}} \right)^2 
\ls 8 \cdot 10^{-31}, 
\ee
which has to be compared with the experimental limit 
of $3 \times 10^{-9}$ \cite{ma}. 
Even worse numbers hold for the other 
elements of the mass matrices and the respective processes they govern, 
see \cite{ich} for a compilation. 
The closer the measured \tm{} value gets to the minimal value (0.1205 for 
our specific example) the smaller gets the allowed 
area or curve in $\alpha$--$\beta$ space and therefore the resulting 
range for the other matrix elements. This minimal value of \tm{} 
corresponds to the $(+--)$ configuration. 
Of course, also for the maximal value of $\tm = 1$ only one pair of phases 
is responsible, however, then holds $m_0 \le 0.2$ eV\@. 
For a precision of 10 $\%$ in the measurement of the minimal  
\tm{} we get a variation in $m_{\mu\mu}$ 
within a factor of 2. A slightly smaller number is obtained for a 
5 $\%$ precision. In Fig.\ \ref{sametm} the area for $\tm = 0.1205 $
is obtained for a precision of 5 $\%$. 
However, it seems to be questionable that these 
values of \tm{} and the required precision are feasible, 
especially in the light of the different results of different nuclear 
matrix element calculations. 
Thus one needs to make compromises in terms of 
theoretical assumptions \cite{matass} 
in order to get the complete mass matrix. 
However, despite the large uncertainty in the other mass matrix entries,   
note that in most cases the mass spectrum of the eigenvalues 
can still be probed: from Fig.\ 26 of Ref.\ \cite{klap} 
it is seen that --- provided the solar solution is known --- some measured 
values of \meff distinguish different schemes.
The many possibilities for $\alpha$ and $\beta$ can be understood in 
the following way: In a three flavor scheme we have 9 parameters: 
3 masses, 3 angles and 3 phases. Oscillation experiments can give 
two mass squared differences (equivalent to 2 masses),  
all 3 angles and 1 phase. The mass scale might be given by the tritium 
spectrum or from cosmology, 
so that we are left with 2 parameters and one observable, namely \mee. 
Only for minimal \tm{} an unambiguous determination 
of the 2 phases is possible, 
which however requires an extremely precise measurement of this 
quantity. 
The smallness of $|U_{e3}|$ allows a broad range for $\beta$, but for 
the $\mu\mu$ entry $\beta$ contributes with 
$c_2 c_3 e^{i \beta} \simeq 1/\sqrt{2} \, e^{i \beta}$, which can have 
a large effect on the magnitude of $m_{\mu\mu}$.\\ 
To conclude, for the $(+--)$ case the uncertainty in $m_{\mu \mu}$ 
is at least of the 
same order as the one stemming from the solar angles and \tm. 
In general the difference can be up to factors around 20.\\
If we choose a smaller $|U_{e3}|$ then the uncertainty in $\beta$ is 
even larger and our arguments are strengthened. For $m_0$ 
smaller then 0.3 eV, Eq.(\ref{hie}) tells us that 
\meff $\ls m_1$, which is the smallest
eigenvalue, lying at the very end of the planned \meff sensitivity. An
analysis of the phases is then probably impossible and again our 
arguments are strengthened.

\section{\label{funf}Conclusions}
\begin{figure}[ht]
\setlength{\unitlength}{1cm}
\begin{center}
\epsfig{file=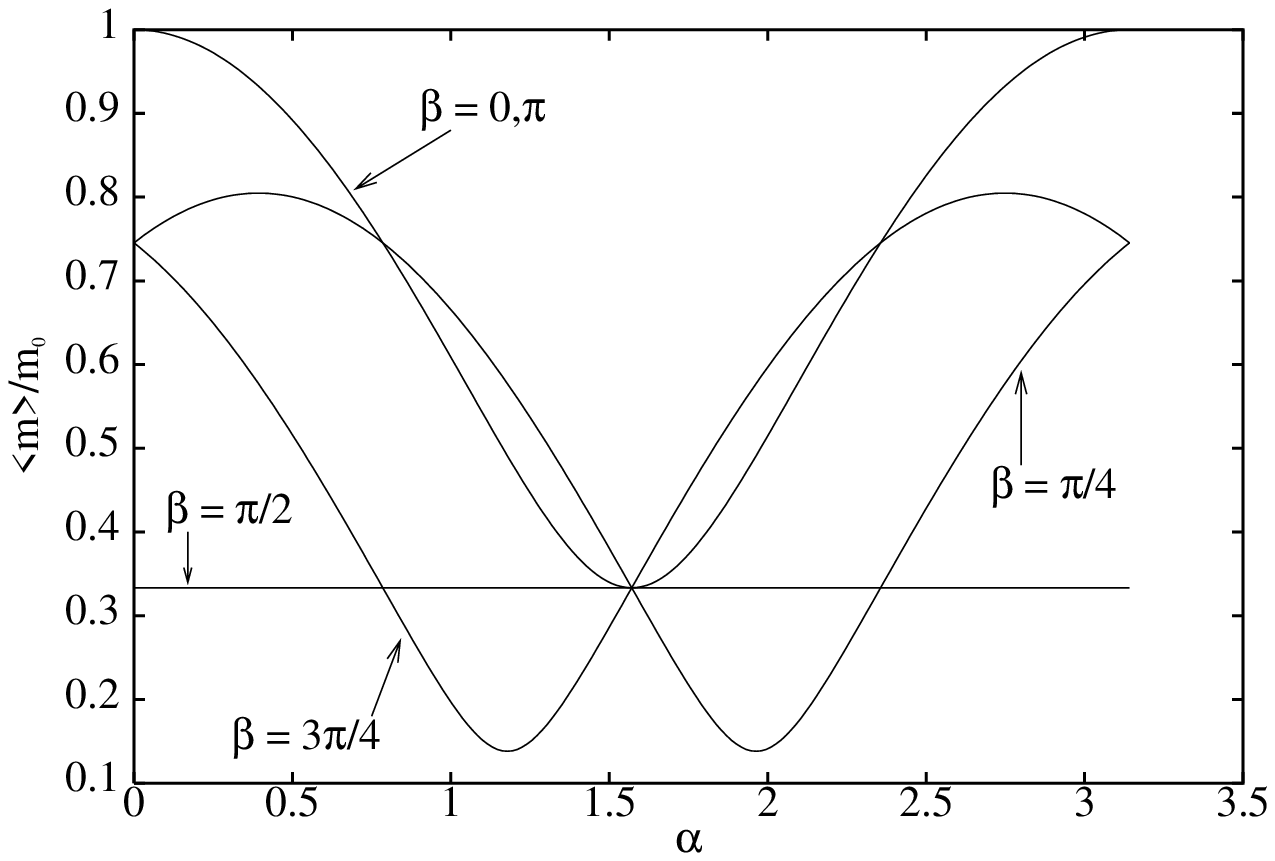,width=13cm,height=8cm}
\end{center}
\caption{\label{triCPVfig}\tm{} for trimaximal mixing 
as a function of one $CP$--violating phase 
for different values of the second phase.}
\begin{center}
\epsfig{file=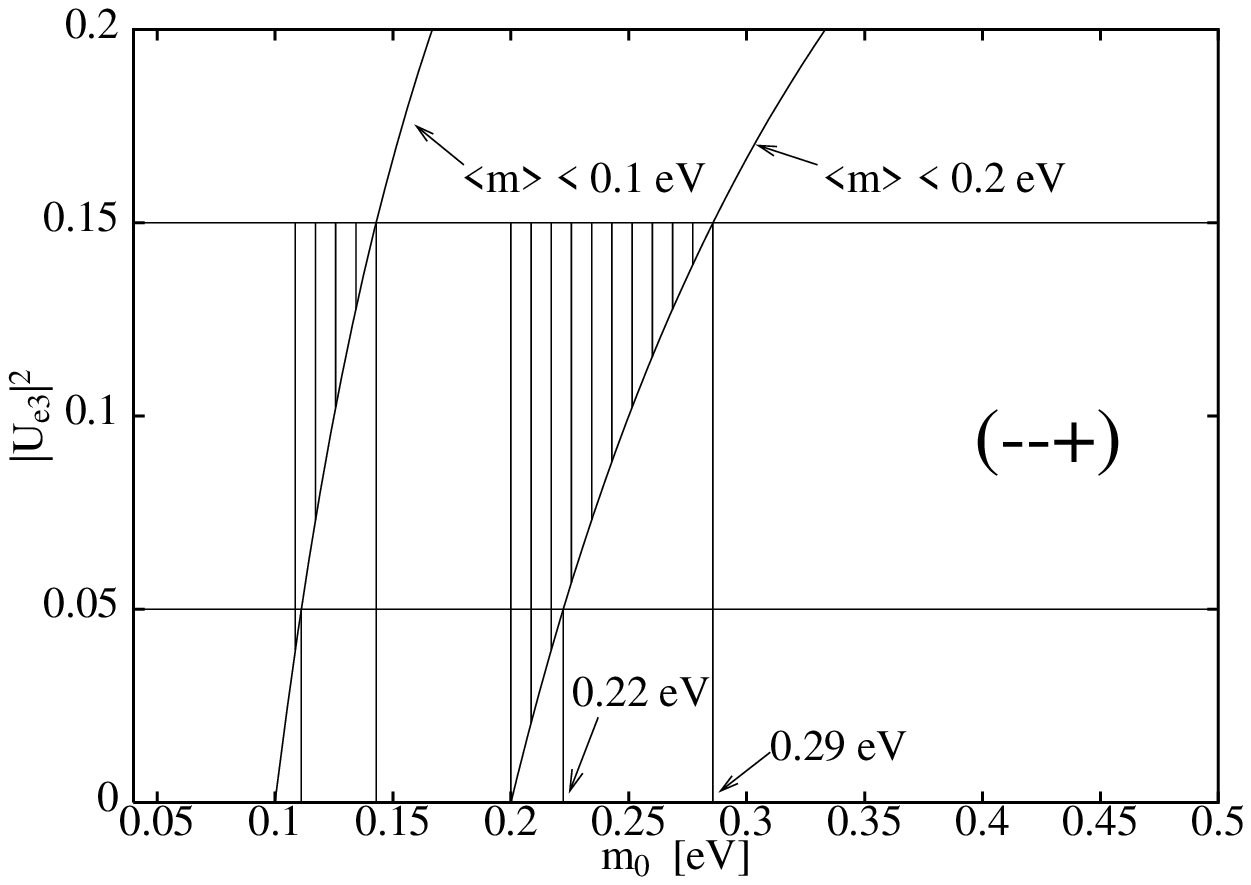,width=13cm,height=7.5cm}
\end{center}
\caption{\label{mmpfig}Allowed range for the common neutrino mass scale $m_0$ 
for $CP$ conservation and the $(--+)$ signature of the $CP$ parities. 
Displayed are the cases $\mee{} \le 0.2$ and 0.1 eV\@. The 
SK`s best fit point corresponds to $m_0 < 0.22 \, (0.11)$ eV and the 
maximal value of $|U_{e3}|^2$ leads to  $m_0 < 0.29 \, (0.14)$ eV for the 
two values of \mee.}
\end{figure}

\begin{figure}[ht]
\setlength{\unitlength}{1cm}
\begin{center}
\epsfig{file=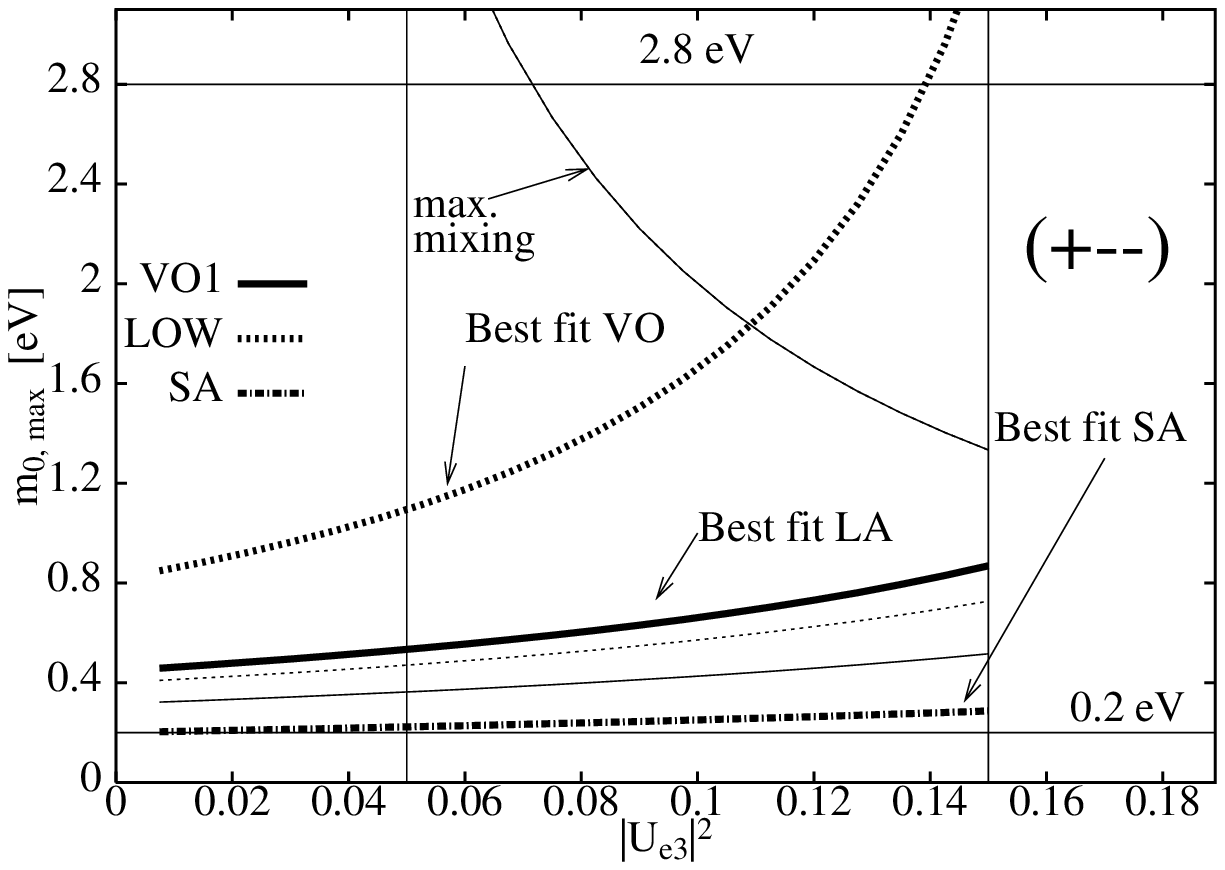,width=13cm,height=7.5cm}
\end{center}
\caption{\label{pmmfig}Maximal $m_0$ as a function of $|U_{e3}|^2$ for 
the VO1, LOW and SA solution and the $(+--)$ signature. 
For LA the plot is very similar to VO1. A limit of $\mee{} \le 0.2$ 
eV is assumed. Allowed is the range under the respective curve.  
The VO2 option of the $(-+-)$ signature is the same as the VO1 option here.}
\begin{center}
\epsfig{file=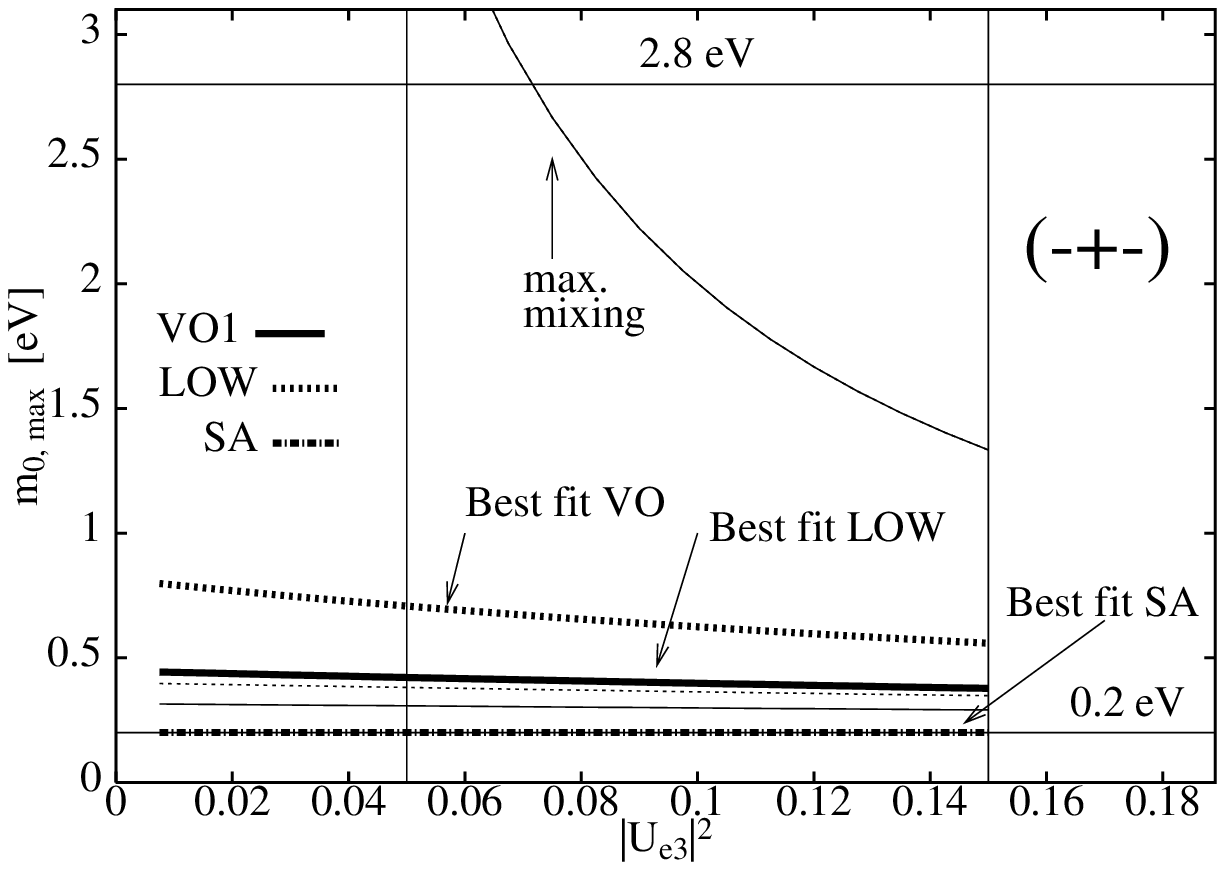,width=13cm,height=8cm}
\end{center}
\caption{\label{mpmfig}Same as the previous figure for the 
$(-+-)$ signature. Again, the 
LA plot is very similar to VO1 and the VO2 option is 
identical to VO1 option in the $(+--)$ case.}
\end{figure}

In a degenerate mass scheme with three neutrinos 
we did a full analysis of the allowed parameter range for the 
relevant function $\tm = \mee{}/m_0$.  
All $CP$--conserving and --violating possibilities for 
cancellation were considered and plots for observables in each case were 
given. When $m_{0, \, \rm max}$ 
is smaller then roughly $0.2 \ldots 0.1$ eV, the limit applies 
no longer to a degenerate scheme but for the highest mass 
of a hierarchical or intermediate scheme. The 
other masses are then given by the measured mass squared differences. 
Then however a signal or improved bound 
from \obb{} is needed to distinguish 
the different possibilities, i.e.\ to verify 
which two masses give which $\Delta m^2$. 
For the SA solution the mass scale is below 0.2 eV\@, 
regardless of $CP$ violation or conservation. Effects of phases 
then lie in the $\%$ range.  
We may summarize the situation for $CP$ invariance such that 
the $(+++)$ case also means $m_0 \le 0.2$ eV\@. 
If the $(--+)$ configuration is realized (Fig.\ \ref{mmpfig}), the 
total mass scale is too small to be directly measured. 
Regarding the other two parity configurations 
(Figs.\ \ref{pmmfig} and \ref{mpmfig}), maximal solar 
mixing allows a broad range for $m_0$, for small $|U_{e3}|^2$ 
even up to the tritium limit of 2.8 eV\@. The smaller the 
angle $\sin^2 2 \theta_{\odot}$ is, the smaller is the maximal allowed 
$m_0$ or the larger is the minimal value of \tm. 
For $CP$ violation two phases are present, in principle varying between 
0 and $\pi$. Because of the smallness of $|U_{e3}|^2 $ the 
dependence on one phase is small. For maximal solar mixing 
complete cancellation and therefore $m_0$ up to 2.8 eV 
is possible. 
From $\sin^2 2 \theta_{\odot} \simeq 0.7$ on, $m_{0, \, \rm max}$ 
approaches 0.2 eV and the degenerate scheme is jeopardized. 
A key value for \tm{} of 0.6 was identified, which would prove the 
existence of $CP$ violation or would rule out some parity configurations. 
This value can unfortunately only be reached if we loosen the 
\meff bound with respect to the uncertainty in nuclear matrix element 
calculations. If we change the limit on \meff by a factor of $x$ 
then all the maximal allowed values for $m_0$ obtained here have to 
be multiplied with this very factor $x$.\\ 
Provided neutrinos are Majorana particles, 
pushing \tm{} below 0.24 would rule out the LA and VO solution, 
and further reduction below 0.06 would rule out also the LOW case. 
If luckily the minimal \tm{} would be measured, an 
uncertainty of 10 $\%$ in \tm{} translates into a variation 
of the other mass matrix entries of about a factor of 2. 
For non--minimal values many choices of the phases are possible. 
This will reflect on the result of the 
other entries in the mass matrices. 
The resulting uncertainty is as least as large as the one resulting from 
measurements of the solar angle and different calculations of nuclear 
matrix elements. 
Because the 
processes governed by these elements are way beyond experimental 
access, further input from the theoretical side is needed. 
Thus, with currently planned projects, experimental verification 
of a given mass matrix is very questionable.

\hspace{3cm}
\begin{center}
{\bf \large Acknowledgments}
\end{center}
I thank A.\ Joshipura, E.\ A.\ Paschos, W.\ G.\ Scott and 
L.\ Wolfenstein 
for helpful comments. 
This work has been supported in part by the
``Bundesministerium f\"ur Bildung, Wissenschaft, Forschung und
Technologie'', Bonn under contract No. 05HT9PEA5.
Financial support from the Graduate College
``Erzeugung und Zerf$\ddot{\rm a}$lle von Elementarteilchen''
at Dortmund university is gratefully acknowledged.

\newpage

\begin{table}[ht]
\begin{center}
\begin{tabular}{c|c|c}
Solution & $\sin^2 2 \theta_\odot$ (90$\%$ C.L.) & Best--fit point \\ \hline
VO  \cite{Inder} & $0.6 \ldots 1 $ & 0.80 \\ \hline 
SA \cite{Spanier} & $7 \cdot 10^{-4} \ldots 10^{-2} $ 
& $5.5 \cdot 10^{-3}$\\ \hline
LA \cite{Spanier} & $0.55 \ldots 1 $ & 0.79 \\ \hline
LOW \cite{Spanier} & $0.75 \ldots 1 $ & 0.94 \\ \hline
\end{tabular}
\caption{\label{data}Solution to the solar neutrino problem, 
90 $\%$ C. L.\ range of the mixing angle and best--fit point of the analysis.}
\end{center}
\end{table}

\newpage

\clearpage

\end{document}